\documentclass[%
 twocolumn,
 amsmath,amssymb,
 aps,prx,
 superscriptaddress,
 tightenlines,
]{revtex4-2}

\usepackage{xcolor}
\usepackage{graphicx}
\usepackage{dcolumn}
\usepackage{bm}

\usepackage{threeparttable}
\usepackage{multirow}
\usepackage{booktabs}
\usepackage{float}
\makeatletter
\let\newfloat\newfloat@ltx
\makeatother
\usepackage{algorithm}
\usepackage{algpseudocode}

\newif\ifprovisional
\provisionaltrue

\begin{document}
\preprint{APS/123-QED}

\title{CLEAR: A Contrastive-Masked Foundation Model for BESIII Event Understanding}
\author{Jingde Chen}
\email{chenjd@ihep.ac.cn}
\affiliation{Institute of High Energy Physics, Chinese Academy of Sciences,
19B Yuquan Road, Shijingshan District, Beijing 100049, China}
\affiliation{University of Chinese Academy of Sciences,
19A Yuquan Road, Shijingshan District, Beijing 100049, China}

\author{Zijie Shang}
\affiliation{Lanzhou University, No. 222 South Tianshui Road, Chengguan District, Lanzhou City, Gansu Province, China}

\author{Tong Liu}
\affiliation{Institute of High Energy Physics, Chinese Academy of Sciences,
19B Yuquan Road, Shijingshan District, Beijing 100049, China}
\affiliation{University of Chinese Academy of Sciences,
19A Yuquan Road, Shijingshan District, Beijing 100049, China}

\author{Yao Zhang}
\affiliation{Institute of High Energy Physics, Chinese Academy of Sciences,
19B Yuquan Road, Shijingshan District, Beijing 100049, China}
\affiliation{University of Chinese Academy of Sciences,
19A Yuquan Road, Shijingshan District, Beijing 100049, China}

\author{Beijiang Liu}
\affiliation{Institute of High Energy Physics, Chinese Academy of Sciences,
19B Yuquan Road, Shijingshan District, Beijing 100049, China}
\affiliation{University of Chinese Academy of Sciences,
19A Yuquan Road, Shijingshan District, Beijing 100049, China}

\author{Changzheng Yuan}
\affiliation{Institute of High Energy Physics, Chinese Academy of Sciences,
19B Yuquan Road, Shijingshan District, Beijing 100049, China}
\affiliation{University of Chinese Academy of Sciences,
19A Yuquan Road, Shijingshan District, Beijing 100049, China}

\author{Xiongfei Wang}
\affiliation{Lanzhou University, No. 222 South Tianshui Road, Chengguan District, Lanzhou City, Gansu Province, China}

\author{Ke Li}
\email{like@ihep.ac.cn}
\affiliation{Institute of High Energy Physics, Chinese Academy of Sciences,
19B Yuquan Road, Shijingshan District, Beijing 100049, China}
\affiliation{University of Chinese Academy of Sciences,
19A Yuquan Road, Shijingshan District, Beijing 100049, China}

\date{\today}

\begin{abstract}
The increasing diversity and complexity of \textcolor{black}{decay channels at high energy physics experiments} pose a challenge to analysis workflows, which require time-consuming, labor-intensive channel-specific selections and optimization.
This motivates the development of foundation models capable of learning a unified representation transferable across different channels.
In this paper, we introduce CLEAR (Contrastive Learning for Event Attention-based Representation), a foundation model designed for physics analysis at BESIII that combines masked modeling and contrastive learning to support decay channel classification and anomaly detection.
Masked modeling trains the encoder to predict masked particle features by learning inter-particle correlations, while contrastive learning improves the local clusterability and linear separability of the representation space for discrimination.
By further examining the balance between alignment and uniformity, we investigate how contrastive learning shapes the local neighborhood geometry of the representation space and how this geometry relates to classification and anomaly detection performance.
Decay channel classification favors a balance between alignment and uniformity, whereas anomaly detection benefits from a greater relative weight on uniformity. 
For exclusive final-state identification, CLEAR achieves higher signal efficiency than conventional cut-based methods at matched purity and provides a framework that can potentially be extended to other high-energy physics experiments.
\end{abstract}

\maketitle

\section{Introduction}
\textcolor{black}{Conventional event selection in particle physics typically relies on channel-specific criteria and optimization, requiring substantial time and specialized experience.
\textcolor{black}{The Beijing Spectrometer III (BESIII) detector~\cite{BESIIIDetector} records $e^+e^-$ collisions at the Beijing Electron--Positron Collider II (BEPCII)~\cite{BEPCII}.
The BESIII physics program spans a wide range of decay channels in the $\tau$--charm energy region~\cite{BESIII,BESIIIYellowBook,BESIIIWhitePaper}.
Differences in final-state composition and intermediate resonance structure call for different particle selection criteria and kinematic constraints~\cite{BESIII-2,BESIII-3}.
Machine learning methods have been applied to object identification and event classification at BESIII~\cite{BESIII-GNN} with Monte Carlo (MC) samples~\cite{BESIIIInclusiveMC,EvtGen,BesEvtGen,Geant4,BESIIIBOSS} and other experiments~\cite{ML_1, ML_2} to improve signal--background discrimination.}
However, these models are generally trained for specific decay channels or analysis tasks, and their application to other processes often requires adaptation and retraining.
These limitations motivate the development of a unified event representation that captures inter-particle correlations, distinguishes different physics processes and can be transferred across decay channel analyses.}

\textit{Foundation models} (FMs) offer a promising framework for pursuing this goal~\cite{Opportunities-Risks-Foundation-Models}.
\textcolor{black}{Through large-scale pretraining, FMs learn general representations that can be adapted to multiple downstream tasks, reducing the need to train a separate representation model from scratch for each analysis.}
\textcolor{black}{Existing FM studies in particle physics have explored pretraining approaches, particularly for jets~\cite{OmniJet, MPM, MPM_2, contrastive_FM}. 
These studies provide a methodological basis for exploring unified event representations at BESIII.}

Two widely used pretraining strategies are masked modeling~\cite{BERT, MAE} and contrastive learning~\cite{SimCLR, DINO, MSE}.
In particle physics, masked modeling typically involves masking a subset of particle constituents within an event or jet and reconstructing the masked information from the remaining unmasked constituents~\cite{OmniJet, MPM, MPM_2}.
By capturing inter-particle correlations, this approach enables foundation models to learn informative event-level representations that encode the underlying physical process.
\textcolor{black}{Contrastive learning} shapes the representation space by bringing positive pairs closer together and separating negative pairs~\cite{SimCLR, contrastive_FM}.
For decay channel discrimination, positive pairs can be constructed from events belonging to the same decay channel, while negative pairs are drawn from different channels.
The resulting representations have been shown to be highly effective for downstream discriminative tasks~\cite{SimCLR, DINO}. 
Combining these strategies offers a way to learn inter-particle correlations while organizing event representations for downstream discrimination.

\textcolor{black}{In this work, we introduce CLEAR (Contrastive Learning for Event Attention-based Representation), a foundation model designed for BESIII that combines masked modeling and contrastive learning.}
Our goal is to learn a unified event representation in which events from different physics processes form distinguishable clusters, enabling classification across the full range of physics processes studied at BESIII.
We incorporate physics-informed inductive biases through two complementary mechanisms at the pair and event levels.
A central focus of this work is to investigate how contrastive learning shapes the representation space and affects downstream performance. 
Motivated by the theoretical connection between InfoNCE~\cite{SimCLR} and alignment--uniformity~\cite{Alignment-and-Uniformity}, we conduct a controlled study in which InfoNCE is replaced by an explicit alignment--uniformity objective, allowing us to vary the relative weights of these two geometric properties.
We evaluate the resulting representations on two downstream tasks: classification of 100 distinct $J/\psi$ decay channels and anomaly detection.
Our results show that contrastive learning substantially improves clusterability and linear separability of the representation space, resulting in consistent performance gains across discriminative downstream tasks. 
The alignment--uniformity study further reveals that classification favors a balance between the two properties, whereas anomaly detection benefits from a greater relative weight on uniformity.
CLEAR also achieves higher signal efficiency than conventional cut-based selections at matched purity for several decay channels.
The resulting pretraining framework may be adapted to event-level analyses at other HEP experiments.


\begin{figure*}[t]
\centering
\includegraphics[width=1.00\hsize]{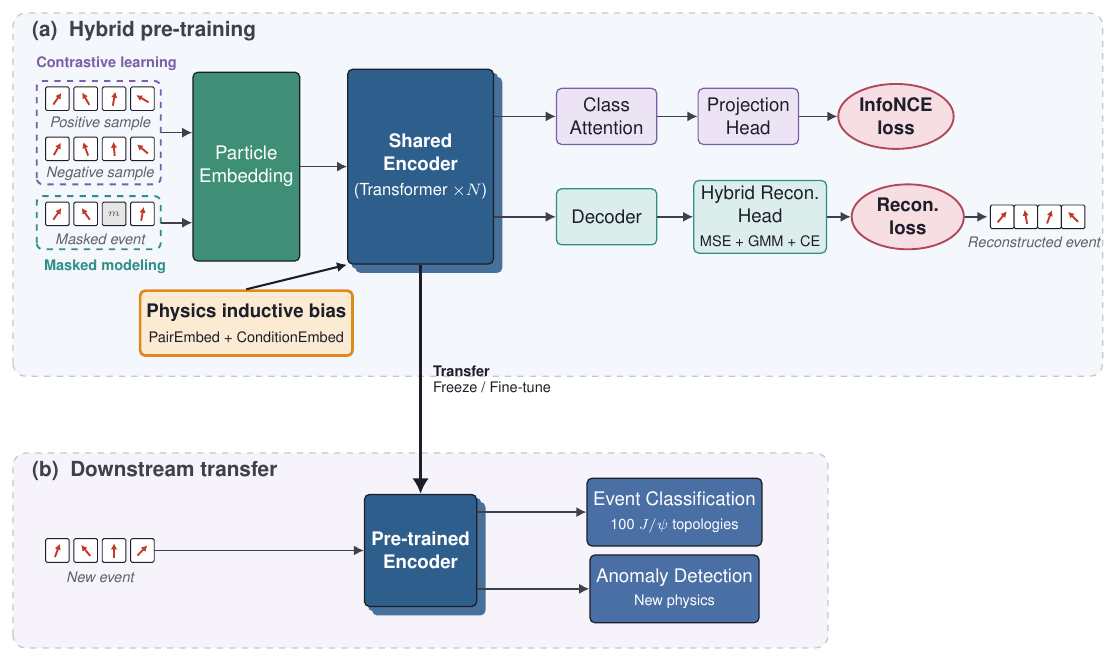}
\caption{Overview of the CLEAR framework.
(a) Hybrid pretraining combines contrastive learning and masked particle modeling within a shared Transformer encoder augmented with physics inductive biases.
(b) The pretrained encoder is transferred to decay-mode classification and anomaly detection through either frozen evaluation or end-to-end fine-tuning.
}
\label{fig:architecture}
\end{figure*}

\section{Related Work}
\textcolor{black}{Deep learning has been extensively applied in particle physics, particularly for jet tagging, with early approaches progressing from jet-image CNNs and tree networks to graph and particle-cloud architectures such as ParticleNet~\cite{deOliveira2016,Guest2016,Louppe2019,Henrion2017,ParticleNet}. 
One key development was the ParticleTransformer (ParT), which represents a jet as a set of constituent particles and incorporates pairwise interaction features as biases in multi-head self-attention mechanism~\cite{ParT}. 
ATLAS deployed the GN2 and its successor GN3~\cite{ATLASGN2,ATLASGN3}.} for flavour tagging, which combines end-to-end processing of low-level jet and track information with physics-informed auxiliary objectives.
Despite these achievements, these models remain inherently task-specific, with each tagging objective requiring a dedicated model trained from scratch on task-specific labeled data. 

This limitation has motivated a shift toward pretrained a FM in HEP, which aim to learn unified representations from large-scale data that transfer across multiple downstream tasks. 
Currently, FMs for HEP have primarily followed two complementary pretraining paradigms.
Masked modeling, as adopted by FMs such as OmniJet~\cite{OmniJet} and Masked Particle Modeling (MPM) learns transferable representations by recovering masked particle information~\cite{MPM, MPM_2}.
Contrastive approaches such as JetCLR instead learn representations by aligning physics-preserving augmented views of the same jet~\cite{Dillon2021}.
\textcolor{black}{EveNet recently extended pretraining to complete collision events, combining masked diffusion-based reconstruction with physics-informed supervised objectives in a shared event representation~\cite{EveNet}}.
These studies explore different pretraining objectives and input structures, providing frameworks for representation learning at both the jet and event levels.

\textcolor{black}{CLEAR is designed to learn event representations for all the decay channels involving different intermediate resonances or final states at BESIII.
Its contrastive objective treats distinct events from the same decay channel as positive pairs and is combined with masked particle modeling.
We further investigate how the relative balance between alignment and uniformity shapes the learned representation and its performance in classification and anomaly detection.}

\section{Model Architecture}
In this section, we introduce CLEAR, a foundation model architecture for the BESIII experiment. The model is designed to learn the unified representations from the inherently heterogeneous and variable-cardinality particle collision data. As shown in Fig.~\ref{fig:architecture}, CLEAR adopts a Siamese framework~\cite{MSE} built on a Transformer backbone enriched with physics-inspired inductive biases~\cite{ParT}. The overall architecture is described in the following subsections.

\subsection{Pretext Tasks}
Our pretraining strategy integrates two complementary pretext tasks: masked modeling and contrastive learning. 

Masked modeling reconstructs the kinematic and detector-level features of a masked particle from the \textcolor{black}{inter-particle correlations} with unmasked particles in the same decay event. 
This method encourages the model to capture the kinematic constraints and conservation patterns during the decay process. 
Unlike natural language tokens, particles in HEP events form permutation-invariant sets without a fixed ordering, making standard positional encodings inapplicable in this case. 
Therefore, we design a specific masking strategy for $J/\psi$ decay channels: rather than masking a fixed proportion of particle information, the proposed \textit{single particle masking strategy} masks the complete feature set of one valid particle in each event, as detailed in Appendix~\ref{app:single-particle-masking}.
This strategy encourages the model to learn the kinematic and topological relationships between the masked particle and the remainder of the event.

The second pretext task is contrastive learning, which aims to shape the \textcolor{black}{global representation space} by bringing the same decay channel closer, while repelling \textcolor{black}{different decay channels}. 
For each anchor sample, we randomly select a positive sample from a pre-defined answer pool using the same Monte Carlo truth decay channel label, while treating the remaining decay channels in the batch as negative samples. 
Although truth labels are currently used to construct positive samples, we retain the standard InfoNCE objective~\cite{SimCLR} rather than adopting a fully supervised contrastive loss~\cite{SCL}, because future versions of CLEAR are intended to use non-label-defined positive samples for fully self-supervised contrastive learning.
InfoNCE loss function can couples into two geometric effects: alignment and uniformity~\cite{Alignment-and-Uniformity}.
To disentangle these effects, we further replace InfoNCE with the explicit alignment--uniformity objective defined in Eq.~\ref{eq:alignment_uniformity}, where $\lambda$ controls the relative emphasis on uniformity over alignment.

\subsection{Pretraining Data and Data Preprocessing}
\label{sec:pretraining-data}
The pretraining data includes \textcolor{black}{100 $J/\psi$ decay channels} ranked by event count in the inclusive Monte Carlo sample~\cite{BESIIIInclusiveMC}.
\textcolor{black}{Particle candidates are reconstructed using the BESIII Offline Software System (BOSS)~\cite{BESIIIBOSS}.
The main drift chamber (MDC) provides charged-track momenta and vertex parameters, while MDC energy-loss measurements and time-of-flight (TOF) information contribute to particle identification~\cite{BESIIIDetector}.
CLEAR also uses shower observables from the electromagnetic calorimeter (EMC) and penetration depth from the muon counter (MUC).}
Because the number of reconstructed particle multiplicity varies across decay event \textcolor{black}{due to} detector response fluctuations and noise, each input is represented as a fixed-size set of $N_{\max}=20$ particle candidates. 
The number of reconstructed particles fewer than 20 in one decay channel are zero-padded, with an attention mask used to distinguish valid particle information from padding fake zeros. 
The number of reconstructed particles more than 20 in one decay channel are truncated to the first 20 reconstructed candidates. 
Each particle data is described by 19 input features, comprising 16 continuous and 3 discrete variables, summarized in Table~\ref{tab:particle-features}.

\begin{table}[t]
\centering
\caption{Particle-level input features used in CLEAR, listed by their original names in the dataset. Crystal-hit counts are treated as continuous inputs,
while EMC timing is encoded categorically.}
\label{tab:particle-features}
\small
\setlength{\tabcolsep}{6pt}
\setlength{\arrayrulewidth}{0.35pt}
\renewcommand{\arraystretch}{1.10}
\begin{tabular}{ll}
\toprule
Feature group & Variables \\
\midrule
\multicolumn{2}{l}{Continuous features (16)} \\
MDC momentum
& \texttt{mdc\_p3\_px}, \texttt{mdc\_p3\_py} \\
& \texttt{mdc\_p3\_pz} \\
Vertex parameters
& \texttt{vz0}, \texttt{vr0} \\
EMC energy
& \texttt{emc\_energy} \\
& \texttt{emc\_e3x3}, \texttt{emc\_e5x5} \\
EMC position
& \texttt{emc\_x}, \texttt{emc\_y}, \texttt{emc\_z} \\
EMC shower shape
& \texttt{emc\_numHits} \\
& \texttt{emc\_secmom}, \texttt{emc\_latmom} \\
Energy--momentum ratio
& \texttt{EoP} \\
MUC response
& \texttt{muc\_dpt} \\
\midrule
\multicolumn{2}{l}{Discrete features (3)} \\
Charge & \texttt{q} \\
Particle identification & \texttt{pid\_type} \\
EMC timing & \texttt{emc\_time} \\
\bottomrule
\end{tabular}
\end{table}

A feature validity mask $\mathbf{M}\in\{0,1\}^{N_{\max}\times19}$ was applied for identifying structurally absent detector responses.
For neutral candidates ($q=0$), charged-particle-specific features are marked as invalid such as MDC momentum. 
EMC features are marked as invalid when both \texttt{emc\_energy} and \texttt{emc\_z} are zero, which is used to identify an absent EMC response. 
Also, a separate particle-level attention mask identifies zero-padded candidates.

Continuous features with approximately Gaussian distributions are standardized directly using $z$-score normalization.
Selected continuous features with non-Gaussian distributions are first mapped to approximately Gaussian distributions using a quantile transformation and are subsequently standardized using $z$-score normalization.
The three discrete features are mapped to categorical indices and encoded using separate 16-dimensional learnable embeddings.
These embeddings are concatenated with the transformed continuous features to construct the particle-level input representation.
The reconstruction results adopted for those different features are described in Section~\ref{sec:loss}

\begin{figure}[t]
\centering
\includegraphics[width=\columnwidth]{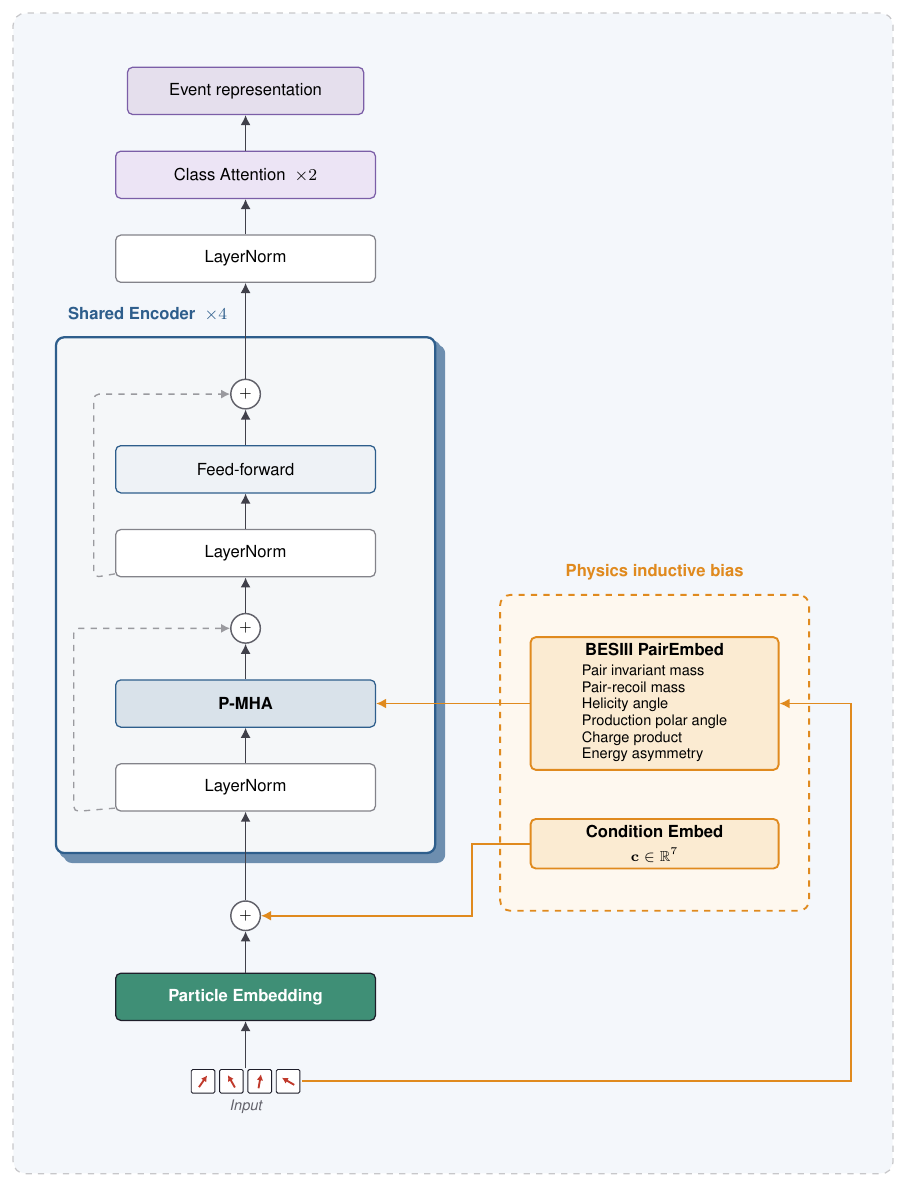}
\caption{Architecture of the CLEAR encoder with event-level conditioning,
pairwise attention biases, and class-attention aggregation.}
\label{fig:encoder}
\end{figure}

\subsection{Encoder}
\label{sec:encoder}
As illustrated in Fig.~\ref{fig:encoder}, the shared encoder consists of standard four pre-norm Transformer blocks~\cite{pre-norm} with an embedding dimension of $D=256$ followed by a final LayerNorm block.
In the contrastive branch and during inference, two class-attention blocks aggregate the encoder outputs of the unmasked input into an event representation, as shown in Fig.~\ref{fig:encoder}.
Prior physics information is incorporated through two complementary pathways: event-level conditioning, which adds an event-condition embedding to the input \texttt{[CLS]} token and pair-level conditioning, which introduces a pairwise interaction bias into particle multi-head attention (P-MHA) block.
These provide global event-level features and local pair-level kinematic relations respectively.

%
%
\subsubsection{Event-Level Embedding}
\label{sec:condition}
Event-level conditioning complements particle-level inputs with information about detector response and event-level kinematic closure.

The event-level conditioning of detector response comprises four variables: the multiplicities of charged and neutral candidates, the number of charged candidates without EMC response and the number of neutral candidates without EMC response.
Even within the same decay channel, these variables can vary because of differences in particle kinematics and detector response.
Those event-level conditioning is intended to help the encoder explicitly account for such reconstruction variations.

The remaining three variables are the signed event-level recoil mass $M_{\mathrm{rec}}^{\mathrm{evet}}$, missing energy $E_{\mathrm{miss}}$ and missing momentum $|\vec p_{\mathrm{miss}}|$, defined as: 
\begin{align}
E_{\mathrm{miss}}
  &= \sqrt{s}-E_{\mathrm{vis}},\\
\vec p_{\mathrm{miss}}
  &= -\vec p_{\mathrm{vis}},\\
M_{\mathrm{rec}}^2
  &= E_{\mathrm{miss}}^2-|\vec p_{\mathrm{miss}}|^2,\\
M_{\mathrm{rec}}^{\mathrm{evet}}
  &= \operatorname{sgn}(M_{\mathrm{rec}}^2)
     \sqrt{|M_{\mathrm{rec}}^2|},
\end{align}
where $\sqrt{s}$ denotes the nominal center-of-mass energy, and $(E_{\mathrm{vis}},\vec p_{\mathrm{vis}})=\sum_i p_i$ is the total reconstructed four-momentum. 
The signed function of $M_{\mathrm{rec}}^{\mathrm{event}}$ preserves negative reconstructed mass-squared values arising from detector resolution.
The energies of charged candidates are calculated from their measured momenta using the masses assigned by particle identification categories, whereas neutral candidates are treated as massless, with their energies given by the EMC energy deposits and their directions defined by the EMC cluster positions.

These three variables provide complementary information for distinguishing decay channels with different multiplicities of neutral particles such as $\pi^0$.
When their decay products are not fully reconstructed, the resulting energy and momentum imbalance provides additional kinematic information to help the model to characterize the event and identify its decay channel.
The resulting event-level conditioning is linearly projected to the embedding and added to the input \texttt{[CLS]} token, denoted Condition Embed as illustrated in Fig.~\ref{fig:encoder}.

%
%
\subsubsection{Pair-Level Embedding}
\label{sec:pair_embed}
Following the P-MHA formulation of the Particle Transformer~\cite{ParT}, we add a learned pairwise interaction bias to the attention logits before the softmax operation.
To describe two-particle relations in BESIII physics, we construct this bias from six pairwise observables: pair invariant mass, pair-recoil mass, helicity angle, production polar angle, charge product and energy asymmetry.
These observables are defined as follows:
\begin{equation}
\begin{aligned}
m_{ij}^{2}
  &= (E_i+E_j)^2-|\vec p_i+\vec p_j|^2,\\
(M_{\mathrm{rec},ij}^{\mathrm{pair}})^2
  &= (\sqrt{s}-E_i-E_j)^2
     -|\vec p_i+\vec p_j|^2,\\
\cos\theta_{\mathrm{hel},ij}
  &= \frac{\vec p_r^{\,*}\cdot\hat n_{ij}}
           {|\vec p_r^{\,*}|},\\
\cos\theta_{ij}^{*}
  &= \frac{p_{i,z}+p_{j,z}}{|\vec p_i+\vec p_j|},\\
Q_{ij}
  &= q_iq_j,\\
|A_{ij}|
  &= \frac{|E_i-E_j|}{E_i+E_j},
\end{aligned}
\label{eq:pair_observables}
\end{equation}
where $E_i$, $\vec p_i$, and $q_i$ denote the reconstructed energy, three-momentum, and charge of candidate and $\sqrt{s}$ is the nominal center-of-mass energy.
For the helicity angle, $\vec p_r^{\,*}$ is the reference candidate's momentum in the pair rest frame, and $\hat n_{ij}=(\vec p_i+\vec p_j)/|\vec p_i+\vec p_j|$ is the pair flight direction.
The reference candidate is selected by larger charge, then larger assigned mass, and finally lower candidate index.
The production polar angle $\theta_{ij}^{*}$ is measured relative to the beam axis.

This block denoted BESIII PairEmbed forms the upper orange pathway in Fig.~\ref{fig:encoder}.
Pairs involving padding and \texttt{[CLS]} receive zero bias.

%
%
\subsubsection{Encoder Blocks and Class Attention}
\label{sec:class_attention}
As shown in Fig.~\ref{fig:encoder}, the particle embeddings and the condition-enhanced \texttt{[CLS]} token are processed by pre-norm Transformer blocks, each consisting of P-MHA and feed-forward networks with residual connections. 
A padding mask excludes padded candidates from attention and a final LayerNorm produces the encoded token representations.
Two class attention blocks based on CaiT~\cite{cait} then aggregate these representations by using \texttt{[CLS]} token as the sole query and the concatenation of the \texttt{[CLS]} and encoded particle tokens as the keys and values.
Only \texttt{[CLS]} token is updated during aggregation and no pairwise interaction bias is applied.
The resulting event-level representation is used for contrastive pretraining and downstream decay channels classification and anomaly detection.

%
%
\subsection{Pretraining Loss Function}
\label{sec:loss}
CLEAR is jointly pretrained by masked modeling and contrastive learning:
\begin{equation}
\mathcal L = w_{\mathrm{Mask}}\mathcal L_{\mathrm{Mask}} + w_{\mathrm{con}}\mathcal L_{\mathrm{con}},
\label{eq:total}
\end{equation}
where the weights follow a curriculum schedule~\cite{curriculum} that initially emphasizes masked modeling and progressively increases the relative contribution of contrastive learning.

One valid particle is randomly selected and removed from each encoder input.
A Transformer decoder reconstructs its features from the encoded remaining particles' information.
The reconstruction objective combines mean squared error (MSE) for continuous features with approximately Gaussian distributions, Gaussian mixture model of negative log-likelihood for complex non-Gaussian multi-modal distribution such as $x$ and $y$ position of EMC hit and cross-entropy for categorical features such as particle identification categories:
\begin{equation}
\mathcal L_{\mathrm{Mask}} = \mathcal L_{\mathrm{MSE}} + \mathcal L_{\mathrm{GMM}} + \mathcal L_{\mathrm{CE}}.
\end{equation}
A feature validity mask excludes undefined detector responses from the reconstruction targets.
This task encourages CLEAR to capture inter-particle correlations within same decay event.

Contrastive learning shapes event representations by bringing positive pairs closer and pushing negative pairs apart in the representation space.
The class attention output is mapped through a projection head and $\ell_2$-normalized.
For each anchor, a positive example is sampled from a separate sampling pool with the same truth decay channel label.
The baseline uses the InfoNCE objective:
\begin{equation}
\mathcal L_{\mathrm{con}} = -\frac{1}{B}\sum_{i=1}^{B}\log\frac{\exp(\mathbf z_i^a\cdot\mathbf z_i^p/\tau)}{\sum_{j=1}^{B}\exp(\mathbf z_i^a\cdot\mathbf z_j^p/\tau)},
\label{eq:infonce}
\end{equation}
where $B$ is the batch size, $\mathbf z_i^a$ and $\mathbf z_i^p$ are the normalized anchor and positive embeddings, respectively. 
The temperature $\tau$ is set to $0.1$ in this paper.
For each anchor, the other $B-1$ anchors are treated as negatives.

In order to study the geometric effects of contrastive learning, we replace InfoNCE with an explicit alignment--uniformity objective~\cite{Alignment-and-Uniformity}.
Alignment favors encoders that assign similar features to similar samples. 
The alignment loss is defined as follows:
\begin{align}
\ell_{\mathrm{align}} &= \mathbb E_{(x,y)\sim p_{\mathrm{pos}}}\left[\|f(x)-f(y)\|_2^2\right]\nonumber\\
&\approx \frac{1}{B}\sum_{i=1}^{B}\|\mathbf z_i^a-\mathbf z_i^p\|_2^2,
\label{eq:alignment}
\end{align}
where $f$ denotes the mapping from an input event to its normalized projected representation and $p_{\mathrm{pos}}$ is the positive pair distribution.
$\|\cdot\|_2$ denotes the Euclidean norm.

Uniformity prefers a feature distribution that preserves maximal information on the unit hypersphere. Let $\mathcal Z=\{\mathbf z_1,\ldots,\mathbf z_{2B}\}$ denote the embeddings in a batch, with $\mathbf z_i=\mathbf z_i^a$ and $\mathbf z_{B+i}=\mathbf z_i^p$ for $i=1,\ldots,B$.
The uniformity loss is defined as follows:
\begin{align}
\ell_{\mathrm{unif}} &= \log \mathbb E_{x,y\overset{\mathrm{i.i.d.}}{\sim}p_{\mathrm{data}}} \left[ \exp\!\left(-2\|f(x)-f(y)\|_2^2\right) \right] \nonumber\\ 
&\approx \log \!\left[\frac{2}{2B(2B-1)} \sum_{\{\mathbf u,\mathbf v\}\subset\mathcal Z}
\exp\!\left(-2\|\mathbf u-\mathbf v\|_2^2\right)\right],
\label{eq:uniformity}
\end{align}
where $p_{\mathrm{data}}$ denotes the event distribution and $\mathrm{i.i.d.}$ (independent and identically distributed) indicates that $x$ and $y$ are independently sampled from this same distribution.
The batch approximation includes all $B(2B-1)$ unordered pairs of distinct batch entries, covering anchor--anchor, positive--positive, and anchor--positive pairs.

For the alignment--uniformity experiments, the contrastive learing is
\begin{equation}
\mathcal L_{\mathrm{con}}(\lambda) = \frac{\ell_{\mathrm{align}}+\lambda\ell_{\mathrm{unif}}}{1+\lambda}, \qquad \lambda\geq 0,
\label{eq:alignment_uniformity}
\end{equation}
where both loss components are evaluated using their batch approximations. 
The factor $1/(1+\lambda)$ normalizes the two weights to sum to one.
Varying this balance allows us to examine how the local neighborhood geometry relates to downstream tasks like decay channel classification and anomaly detection.
InfoNCE serves as the baseline for the alignment--uniformity experiments on decay channel classification and anomaly detection.

%
%
%
%
\begin{figure*}[t]
\centering
\includegraphics[width=\textwidth]{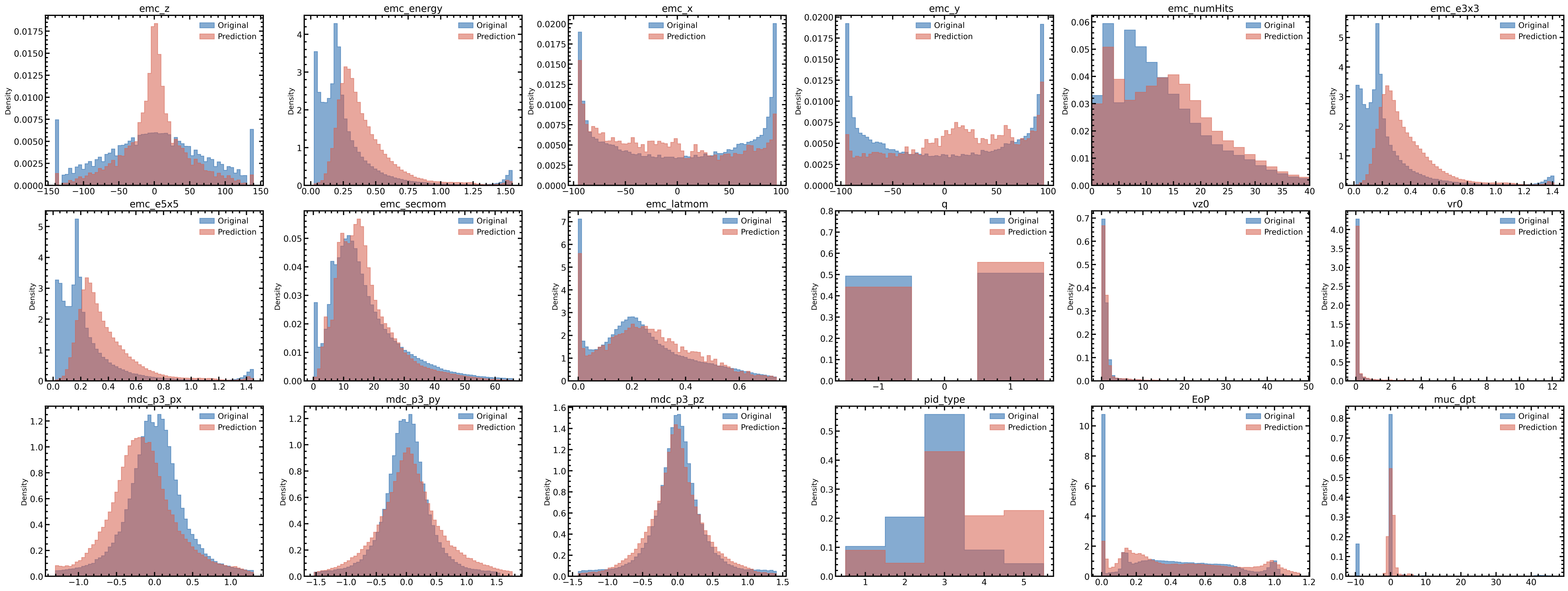}
\caption{
Comparison between the original (blue) and reconstructed (red) distributions of observables in the masked modeling task.
\textcolor{black}{CLEAR reproduces the overall distributions of most continuous observables, while discrepancies remain in peak positions, widths, relative peak heights and particularly particle identification categories.
}}
\label{fig:masked-reconstruction}
\end{figure*}

\section{Results}
\label{sec:results}
We pretrain the CLEAR model using Monte Carlo samples of $J/\psi$ decays including 100 decay channels. 
We evaluate its performance across three tasks: (i) masked particle reconstruction, (ii) decay channels classification, and (iii) anomaly detection. 
Unless otherwise specified, the datasets used for downstream evaluation are entirely separate from the pretraining dataset.

\subsection{Masked Particle Reconstruction}
Following joint pretraining with masked particle modeling and contrastive learning, we evaluate the decoder predictions for masked particle features.
As described in Sec.~\ref{sec:loss}, the reconstruction objective combines MSE for approximately Gaussian continuous features, a GMM-based negative log-likelihood loss for features with more complex non-Gaussian distributions and cross-entropy for categorical features.
\textcolor{black}{Figure~\ref{fig:masked-reconstruction} compares the original and reconstructed distribution. 
CLEAR captures broad distributional characteristics across a range of observables, including the approximately Gaussian shape of \texttt{mdc\_p3\_pz}, the right-skewed profiles of \texttt{emc\_secmom} and \texttt{emc\_latmom}, and the sharp peaks near zero with extended tails in \texttt{vz0} and \texttt{vr0}.
It also approximately reproduces the U-shaped distributions of the $x$ and $y$ coordinates of EMC shower positions.
These results suggest that masked particle modeling enables CLEAR to learn meaningful statistical structure across particle-level observables with diverse distributional shapes. 
Nevertheless, discrepancies remain in the peak structures of some EMC observables, such as $z$ position of EMC hit and are particularly evident in the relative frequencies of particle identification categories.
Overall, these results highlight the potential of CLEAR for masked particle reconstruction and motivate future refinements to the decoder and feature-specific distribution modeling to improve reconstruction fidelity across observables.}

\subsection{Decay Channel Classification}
\label{sec:classification}
We evaluate the effectiveness of the representations learned by CLEAR for downstream decay channel classification. 
To assess performance under the strongly imbalanced decay channel distribution, we report macro-averaged $F_1$, support-weighted $F_1$ and accuracy:
\begin{equation}
\begin{aligned}
F_{1,\mathrm{macro}}
&= \frac{1}{C}\sum_{c=1}^{C}F_{1,c}, \quad F_{1,\mathrm{weighted}} = \sum_{c=1}^{C}\frac{n_c}{N}F_{1,c}, \\
\mathrm{Acc} &= \frac{1}{N}\sum_{i=1}^{N}\mathbb{I}(\hat{y}_i=y_i),
\end{aligned}
\end{equation}
\textcolor{black}{where $C=100$ is the number of decay channels, $n_c$ is the number of test events in channel $c$ and $N=\sum_c n_c$ is the total number of test events.}
The predicted and true channel labels for event $i$ are denoted by $\hat{y}_i$ and $y_i$ and $\mathbb{I}(\cdot)$ is the indicator function.
The per-channel score is defined as $F_{1,c}=2P_cR_c/(P_c+R_c)$, where $P_c=\mathrm{TP}_c/(\mathrm{TP}_c+\mathrm{FP}_c)$ and $R_c=\mathrm{TP}_c/(\mathrm{TP}_c+\mathrm{FN}_c)$ are the precision and recall for channel $c$.
Here, $\mathrm{TP}_c$, $\mathrm{FP}_c$, and $\mathrm{FN}_c$ denote the corresponding true-positive, false-positive and false-negative counts.
Macro-averaged $F_1$ gives equal weight to each decay channel, whereas support-weighted $F_1$ weights each channel by its frequency in the test sample.
All results in this subsection use CLEAR pretrained with InfoNCE as the contrastive objective.
The contribution of contrastive learning and the trade-off between alignment and uniformity are examined separately in Sec.~\ref{sec:ablation}.
We present three complementary analyses: evaluation with a frozen encoder, comparison of fine-tuning strategies and analysis of final-state discrimination.

\subsubsection{Frozen-Encoder Evaluation}
\label{sec:frozen}
In order to evaluate the representations acquired from pretraining, the encoder of the CLEAR model are frozen and only lightweight classifiers are trained using these event representations. 
For the downstream task --- decay channel classification, we are using four lightweight classifiers: a $k$-NN classifier, a linear probe implemented using logistic regression, a two-layer MLP and a three-layer MLP. 
\begin{table}[t]
\centering
\caption{Frozen-encoder evaluation for $J/\psi$ decay channel
classification using four lightweight classifiers.}
\label{tab:frozen}
\begin{tabular}{lccc}
\hline
Head & Acc & $F_{1,\mathrm{macro}}$ & $F_{1,\mathrm{weighted}}$ \\
\hline
$k$-NN               & 64.45\%  & 47.91\% & 62.46\%   \\
Linear               & 68.45\%  & 51.77\% & 65.68\%   \\
MLP (2 layers)       & 69.06\%  & 52.36\% & 66.13\%   \\
MLP (3 layers)       & 68.96\%  & 51.23\% & 65.80\%   \\
\hline
\end{tabular}
\end{table}

As shown in Table~\ref{tab:frozen}, the linear probe achieves an accuracy of $68.45\%$, only $0.61\%$ below the best-performing nonlinear classifier, the two-layer MLP. 
This small difference indicates that the discriminative information relevant to decay channel classification is accessible to a linear classifier in the frozen representation space. 
Increasing the classifier depth from two to three layers yields no further improvement. 
This result suggests that the remaining limitation is associated primarily with the frozen representations rather than with the classifier depth. 
These results motivate unfreezing and fine-tuning the encoder to further improve downstream performance, as investigated in the following subsection.

\subsubsection{Fine-Tuning Strategies}
\label{sec:finetune}
The frozen encoder results show that increasing the layers of the classification head shows the limitation of improvements for $J/\psi$ decay channel classification. 
Therefore, we unfreeze the encoder of CLEAR model and compare four fine-tuning strategies and all strategies employ the same 2L-MLP classification head.
\begin{enumerate}
    \item \textbf{Class-attention-only (CA-only):} The encoder blocks remain frozen, while only the class attention block and the classification head are updated.

    \item \textbf{End-to-end:} The full encoder is optimized using different learning rates, where a smaller learning rate is assigned to the pretrained backbone and a larger learning rate is applied to the classification head.

    \item \textbf{Multitask:} Masked particle reconstruction is further introduced as an auxiliary loss objective. The loss function is formulated to preserve features learned via masked modeling during task‑specific adaptation.

    \item \textbf{Two-stage:} Only the class attention components and classification head are updated in the initial phase, after which the full encoder is unfrozen to perform end‑to‑end fine‑tuning.
\end{enumerate}

\begin{table}[t]
\centering
\caption{Comparison of fine-tuning strategies on the 100 class $J/\psi$ decay channel classification. The best result in each column is highlighted in bold.}
\label{tab:finetune}
\begin{tabular}{lccc}
\hline
Strategy & Acc & $F_{1,\mathrm{macro}}$ & $F_{1,\mathrm{weighted}}$ \\
\hline
2L-MLP (baseline) & 69.06\% & 52.36\% & 66.13\% \\
CA-only & 69.85\% & 53.37\% & 66.95\% \\
End-to-end & 71.91\% & \textbf{57.12\%} & 69.55\% \\
Multitask & \textbf{71.94\%} & 57.02\% & \textbf{69.56\%} \\
Two-stage & 71.65\% & 56.11\% & 68.99\% \\
\hline
\end{tabular}
\end{table}

As shown in Table~\ref{tab:finetune}, CA-only is the only strategy that keeps most of the CLEAR encoder frozen, with approximately 10\% of the model parameters being updated. 
By contrast, the end-to-end, multitask, and two-stage strategies ultimately unfreeze the entire encoder and perform end-to-end adaptation with the different optimization objectives. 
Updating approximately 10\% of the model parameters in the CA-only strategy yields an accuracy improvement of less than 1\%. 
Fully unfreezing the encoder provides a further improvement of approximately 2\%. 
Therefore, the relatively limited improvement obtained through full fine-tuning suggests that the pretrained CLEAR representations already capture substantial discriminative information for decay channel classification.

%
%
\begin{figure*}[t]
\centering
\includegraphics[width=\textwidth]{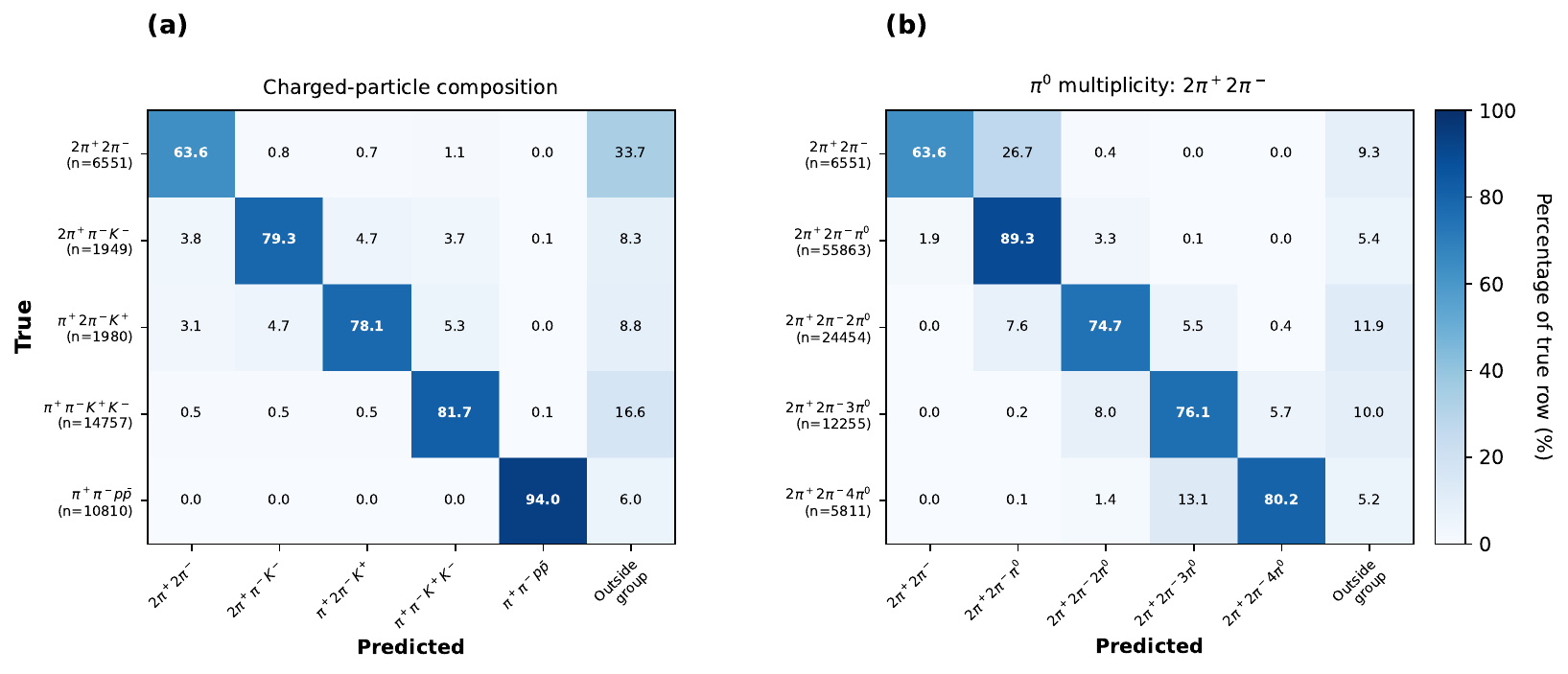}
\caption{
Row-normalized confusion matrices for final-state discrimination: (a) charged-particle composition, (b) $\pi^0$ multiplicity. $n$ denotes the number of test events. ``Outside group'' includes predictions outside the displayed group.
}
\label{fig:final-state}
\end{figure*}

\subsubsection{Final-State Discrimination}
\label{sec:classwise}
\textcolor{black}{We first examine how well CLEAR distinguishes final states with different charged particle compositions.
Figure~\ref{fig:final-state} shows the row-normalized confusion matrices obtained with the encoder frozen.
In each panel, the ``Outside group'' column includes all predictions assigned to final states outside the displayed group. 
Figure~\ref{fig:final-state}(a) compares five final states, each containing four charged particles and no neutral particles.
CLEAR shows little confusion among these compositions: for each true final state, at most $5.3\%$ of events are assigned to any one of the other four.
The $\pi^+\pi^-p\bar{p}$ final state has the highest recall ($94.0\%$).
The $2\pi^+2\pi^-$ final state has the lowest recall ($63.6\%$), yet only about $2.6\%$ of its events are assigned to the other four compositions combined.
A much larger fraction ($33.7\%$) falls outside the displayed group.
To identify where these events are assigned, we next examine final states with the same $2\pi^+2\pi^-$ charged-particle composition and zero to four additional $\pi^0$ mesons.
Figure~\ref{fig:final-state}(b) shows that $26.7\%$ of all true $2\pi^+2\pi^-$ events are classified as $2\pi^+2\pi^-\pi^0$.
This single final state accounts for approximately $79\%$ of the outside-group predictions for true $2\pi^+2\pi^-$ events in panel~(a).
Across the five $\pi^0$ multiplicities, CLEAR identifies the correct final state in $63.6\%$ to $89.3\%$ of events in each category.
For every category, the most frequent misclassification within this family differs from the true final state by just one $\pi^0$.}

%

\subsection{Representation Analysis and Ablation Studies}
\label{sec:ablation}
The downstream results presented in Section~\ref{sec:classification} demonstrate that CLEAR achieves strong performance in decay channel classification in BESIII experiment. 
However, these results do not show how each pretraining component contributes to classification performance and how contrastive learning shapes the geometry of the representation space.
Therefore, we evaluate the pretrained representations using four complementary downstream strategies: $k$-NN, logistic regression, MLP-2L and end-to-end fine-tuning. 
$k$-NN and logistic regression measure clusterability and linear separability of the representation space. 
MLP-2L tests whether a modest nonlinear classifier can separate additional discriminative information from the frozen representations.
Finally, end-to-end fine-tuning jointly optimizes the encoder and classification head, estimating the CLEAR's task-specific performance. 
In this setting, CLEAR functions as the initialization of a large task-specific classifier, thereby estimating CLEAR's practical classification ceiling rather than directly evaluating the quality of its representations.

\subsubsection{Contribution of Contrastive Learning}
\label{sec:contrastive}
\begin{figure*}[t]
\centering
\includegraphics[width=\textwidth]{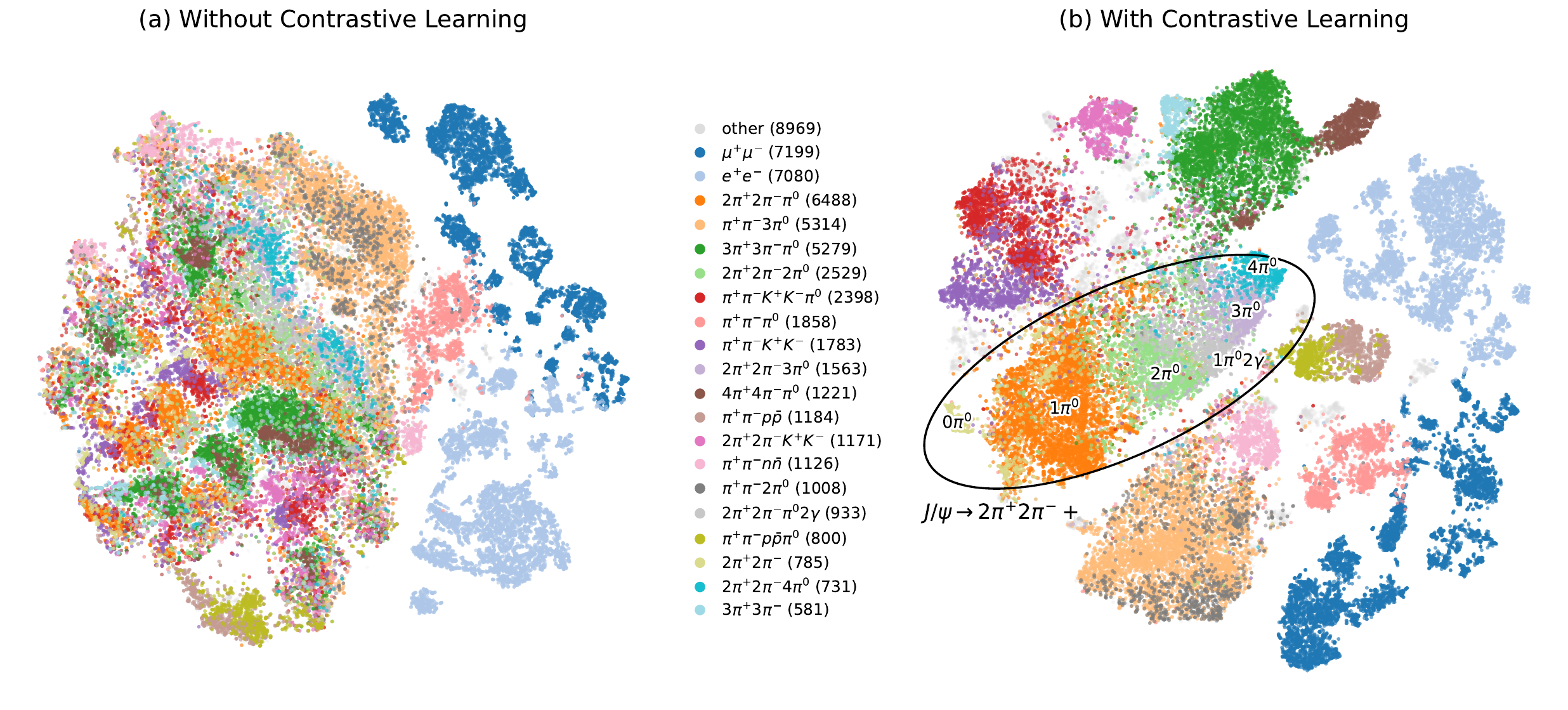}
\caption{
t-SNE visualizations of CLEAR event representations (a) without contrastive learning (Mask-only) and (b) with contrastive learning using InfoNCE (With-CL).
Each point represents a $J/\psi$ decay event, with the top 20 decay channels distinguished by color, representative groups with similar final-state particle compositions outlined in black, and the remaining modes grouped as ``other'' in light gray.
Compared with the Mask-only, With-CL produces tighter and better-separated clusters, resulting in improved $k$-NN and linear-classification accuracies, as reported in Table~\ref{tab:contrastive}.
}
\label{fig:contrastive-representation}
\end{figure*}

To evaluate the contribution of contrastive learning, CLEAR was re-pretrained without the contrastive term, retaining only the masked modeling with reconstruction loss (Mask-only). 
The Mask-only encoder was evaluated under the same experimental settings and downstream protocols as the full model. 
The complete CLEAR model was pretrained using contrastive learning with InfoNCE loss function (With-CL), serves as the baseline.
\begin{table}
\centering
\caption{Contribution of contrastive learning across different downstream
evaluation strategies. Improvements ($\Delta$) are reported in last column.}
\label{tab:contrastive}
\begin{tabular}{lccc}
\toprule
Strategy & Mask-only & With-CL & $\Delta$ \\
\midrule
$k$-NN     & 51.50\% & 64.47\% & $+12.97\%$ \\
Linear     & 61.90\% & 68.44\% & $+6.54\%$  \\
MLP-2L     & 64.62\% & 69.06\% & $+4.44\%$  \\
\midrule
End-to-End & 68.09\% & 71.91\% & $+3.82\%$  \\
\bottomrule
\end{tabular}
\end{table}
As shown in Table~\ref{tab:contrastive}, introducing contrastive learning consistently improves decay channel classification across all downstream strategies. 
The largest improvement is observed for $k$-NN with biggest increasing by 12.97\%, while the linear classifier also improves by 6.54\%. 
As illustrated in Fig.~\ref{fig:contrastive-representation}, contrastive learning substantially enhances both the clusterability and linear separability of the representation space. 
In particular, within the black contour in panel~(b), channels with the same $2\pi^+2\pi^-$ charged-particle composition form an elongated region in which events are arranged in layers according to their $\pi^0$ and $\gamma$ multiplicities.
This results shows that the representation captures differences in neutral $\pi^0$ and $\gamma$ multiplicities within this same charged final state.
The smaller improvement for MLP-2L (4.44\%) and end-to-end fine-tuning (3.82\%) suggests contrastive learning remains beneficial for decay channel classification. 

%
%
\subsubsection{Alignment--Uniformity Trade-off}
\label{sec:lambda}
In order to further evaluate the geometric effects induced by contrastive learning, we replace the InfoNCE objective with the explicit alignment--uniformity formulation defined in Eq.~\ref{eq:alignment_uniformity}. 
We sweep $\lambda \in {0,0.1,0.3,1,2,10}$ from alignment-only to increasingly uniformity-weighted objectives, with the factor $1/(1+\lambda)$, which are limiting changes in the overall loss scale.

\begin{table}[t]
\centering
\caption{
Decay channel classification acaccuracy across alignment--uniformity weights. 
The $\lambda=0$ run is omitted because its representations collapsed. 
Bold values indicate the best result within the alignment--uniformity sweep. 
InfoNCE is the baseline.}
\label{tab:lambda-sweep}
\setlength{\tabcolsep}{6pt}
\setlength{\arrayrulewidth}{0.35pt}
\renewcommand{\arraystretch}{1.10}
\begin{tabular}{lccc|c}
\toprule
Objective & $k$-NN & Linear & MLP-2L & End-to-End \\
\midrule
$\lambda=0.1$ & 58.99\% & 64.60\% & 66.03\% & 68.81\% \\
$\lambda=0.3$ & 61.01\% & 65.91\% & 66.92\% & 69.35\% \\
$\lambda=1$   & \textbf{63.26\%} & \textbf{66.85\%} & \textbf{67.56\%} & \textbf{70.24\%} \\
$\lambda=2$   & 62.85\% & 66.43\% & 67.12\% & 69.96\% \\
$\lambda=10$  & 62.06\% & 65.35\% & 66.12\% & 68.64\% \\
\midrule
InfoNCE       & 64.47\% & 68.44\% & 69.06\% & 71.91\% \\
\bottomrule
\end{tabular}
\end{table}

As shown in Table~\ref{tab:lambda-sweep}, the alignment-only objective ($\lambda=0$) leads to representation collapse, indicating the uniformity term actually prevents representation collapse for CLAER. 
All four downstream strategies achieve their highest accuracy at $\lambda=1$, followed by a decline as $\lambda$ increases. 
Therefore, these results show that uniformity is needed to prevent collapse and that the best decay channel classification performance occurs at a balance between alignment and uniformity.
The $k$-NN accuracy varies by 4.27 percentage points across the stable configurations, compared with at most 2.25 points for the other strategies. 
Because $k$-NN uses distances between pretrained representations directly, this sensitivity indicates that $\lambda$ shapes their local neighborhood geometry.

None of the alignment--uniformity configurations surpasses InfoNCE, which remains ahead of the best configuration for $\lambda=1$ ranging from 1.21 to 1.67 percentage points across the four strategies. 
One possible limitation is that changing the alignment--uniformity balance alone may not adequately resolve decay channels with identical final-state particle compositions but different intermediate resonances.

\subsection{Anomaly Detection}
\label{sec:ad}
The classification task evaluates the discriminative power of representations for 100 classes decay channels in the pretraining data.
Anomaly detection addresses a complementary question: whether the pretrained representation space can distinguish events from unseen decay channels without task-specific anomaly training or labeled signal samples.
To reflect model-independent searches for new physics at BESIII, where signal hypotheses are not predefined, we freeze the pretrained encoder and derive anomaly scores from the geometry of its representation space.
By eliminating a trainable downstream classifier, this design probes how effectively the pretrained representations separate unseen decay channels from the pretraining data.

\subsubsection{Comparison of Anomaly Scorers}
\label{sec:ad-scorers}
We consider two categories of out-of-distribution (OOD) anomalies with distinct physical characteristics: \textit{far-OOD} and \textit{near-OOD}. 
Far-OOD anomalies correspond to decay channels whose final states are absent from the pretraining data, whereas near-OOD anomalies share final-state with pretraining decay channels but with an unseen intermediate resonance.
The near-OOD setting therefore probes sensitivity to previously unseen resonance structures, providing a benchmark relevant to searches for new resonances, particularly important for HEP.

We construct four test sets: Test-1 and Test-2 for near-OOD anomalies, and Test-3 and Test-4 for far-OOD anomalies.
Each set contains 200000 events with no event-level overlap with the pretraining sample. 180000 normal events from the 100 in-distribution decay channels and 20000 anomalous events from two out-distribution decay channels. 
Table~\ref{tab:ad-test-definitions} lists the anomalous decay trees and their final states.
\begin{table*}[t]
\centering
\caption{Anomalous decay channels defining the four anomaly detection test sets.}
\label{tab:ad-test-definitions}
\footnotesize
\setlength{\tabcolsep}{4pt}
\renewcommand{\arraystretch}{1.15}
\begin{tabular}{llcll}
\toprule
Test set & Regime & \shortstack{Channel label}
& Anomalous decay tree & Final state \\
\midrule
Test-1 & near-OOD & N1
& \parbox[t]{0.42\textwidth}{\raggedright
  \(J/\psi\to\pi^{+}\Delta^{+}\bar{\Delta}^{--}\);
  \(\Delta^{+}\to\pi^{0}p\);
  \(\bar{\Delta}^{--}\to\pi^{-}\bar{p}\)\strut}
& \parbox[t]{0.19\textwidth}{\raggedright
  \(\pi^{0}\pi^{+}\pi^{-}p\bar{p}\)\strut} \\

& & N2
& \parbox[t]{0.42\textwidth}{\raggedright
  \(J/\psi\to\pi^{0}\eta b_{1}^{0}\);
  \(\eta\to\pi^{0}\pi^{+}\pi^{-}\);
  \(b_{1}^{0}\to\pi^{0}\omega\);
  \(\omega\to\pi^{0}\pi^{+}\pi^{-}\)\strut}
& \parbox[t]{0.19\textwidth}{\raggedright
  \(4\pi^{0}2\pi^{+}2\pi^{-}\)\strut} \\

\addlinespace[2pt]
\midrule
Test-2 & near-OOD & N3
& \parbox[t]{0.42\textwidth}{\raggedright
  \(J/\psi\to\pi^{0}\eta^{\prime}b_{1}^{0}\);
  \(\eta^{\prime}\to\pi^{+}\pi^{-}\gamma\);
  \(b_{1}^{0}\to\pi^{0}\omega\);
  \(\omega\to\pi^{0}\pi^{+}\pi^{-}\)\strut}
& \parbox[t]{0.19\textwidth}{\raggedright
  \(3\pi^{0}2\pi^{+}2\pi^{-}\gamma\)\strut} \\

& & N4
& \parbox[t]{0.42\textwidth}{\raggedright
  \(J/\psi\to\Delta^{++}\bar{\Delta}^{--}\);
  \(\Delta^{++}\to\pi^{+}p\);
  \(\bar{\Delta}^{--}\to\pi^{-}\bar{p}\)\strut}
& \parbox[t]{0.19\textwidth}{\raggedright
  \(\pi^{+}\pi^{-}p\bar{p}\)\strut} \\

\addlinespace[2pt]
\midrule
Test-3 & far-OOD & F1
& \parbox[t]{0.42\textwidth}{\raggedright
  \(J/\psi\to\pi^{0}p\bar{p}\)\strut}
& \parbox[t]{0.19\textwidth}{\raggedright
  \(\pi^{0}p\bar{p}\)\strut} \\

& & F2
& \parbox[t]{0.42\textwidth}{\raggedright
  \(J/\psi\to\pi^{+}\pi^{-}K^{*+}K^{*-}\);
  \(K^{*+}\to\pi^{+}K^{0}\);
  \(K^{*-}\to\pi^{-}\bar{K}^{0}\);
  \(K^{0}\to K_{L}^{0}\);
  \(\bar{K}^{0}\to K_{L}^{0}\)\strut}
& \parbox[t]{0.19\textwidth}{\raggedright
  \(2K_{L}^{0}2\pi^{+}2\pi^{-}\)\strut} \\

\addlinespace[2pt]
\midrule
Test-4 & far-OOD & F3
& \parbox[t]{0.42\textwidth}{\raggedright
  \(J/\psi\to\bar{K}^{*}K^{*+}b_{1}^{-}\);
  \(\bar{K}^{*}\to\pi^{+}K^{-}\);
  \(K^{*+}\to\pi^{+}K^{0}\);
  \(b_{1}^{-}\to\pi^{-}\omega\);
  \(K^{0}\to K_{S}^{0}\);
  \(\omega\to\pi^{0}\pi^{+}\pi^{-}\);
  \(K_{S}^{0}\to\pi^{+}\pi^{-}\)\strut}
& \parbox[t]{0.19\textwidth}{\raggedright
  \(\pi^{0}4\pi^{+}3\pi^{-}K^{-}\)\strut} \\

& & F4
& \parbox[t]{0.42\textwidth}{\raggedright
  \(J/\psi\to K^{*}K^{*-}b_{1}^{+}\);
  \(K^{*}\to\pi^{-}K^{+}\);
  \(K^{*-}\to\pi^{-}\bar{K}^{0}\);
  \(b_{1}^{+}\to\pi^{+}\omega\);
  \(\bar{K}^{0}\to K_{S}^{0}\);
  \(\omega\to\pi^{0}\pi^{+}\pi^{-}\);
  \(K_{S}^{0}\to\pi^{+}\pi^{-}\)\strut}
& \parbox[t]{0.19\textwidth}{\raggedright
  \(\pi^{0}3\pi^{+}4\pi^{-}K^{+}\)\strut} \\
\bottomrule
\end{tabular}
\end{table*}

All results in Table~\ref{tab:ad-scorers} are obtained using the CLEAR model pretrained with the InfoNCE contrastive objective.
We evaluate five anomaly scorers: Mahalanobis distance, $k$-NN distance, Isolation Forest, reconstruction mean-squared error (Recon-MSE) and an ensemble scorer. 
Detection performance is measured by the area under the receiver operating characteristic curve (AUROC). 
We calculate AUROC separately for the two anomalous decay channels in each test set.
\begin{table*}[t]
\centering
\caption{AUROC for each anomalous decay channel in the four test sets. 
Column headings give testing sample labels. 
Bold entries indicate the highest AUROC for each channel.}
\label{tab:ad-scorers}
\small
\setlength{\tabcolsep}{6pt}
\renewcommand{\arraystretch}{1.15}
\begin{tabular}{lcccccccc}
\toprule
& \multicolumn{4}{c}{Near-OOD}
& \multicolumn{4}{c}{Far-OOD} \\
\cmidrule(lr){2-5}\cmidrule(lr){6-9}
& \multicolumn{2}{c}{Test-1}
& \multicolumn{2}{c}{Test-2}
& \multicolumn{2}{c}{Test-3}
& \multicolumn{2}{c}{Test-4} \\
\cmidrule(lr){2-3}\cmidrule(lr){4-5}
\cmidrule(lr){6-7}\cmidrule(lr){8-9}
Scorer & N1 & N2 & N3 & N4 & F1 & F2 & F3 & F4 \\
\midrule
$k$-NN distance
& 0.69 & 0.69 & 0.59 & 0.51
& 0.83 & \textbf{0.83} & 0.67 & 0.68 \\
Mahalanobis
& \textbf{0.77} & \textbf{0.77}
& \textbf{0.64} & \textbf{0.64}
& 0.89 & 0.78 & \textbf{0.77} & \textbf{0.78} \\
Isolation Forest
& 0.60 & 0.57 & 0.41 & 0.44
& \textbf{0.95} & 0.49 & 0.49 & 0.49 \\
Recon-MSE
& 0.47 & 0.41 & 0.41 & 0.50
& 0.54 & 0.49 & 0.42 & 0.42 \\
Ensemble
& 0.72 & 0.70 & 0.56 & 0.58
& 0.95 & 0.73 & 0.67 & 0.68 \\
\bottomrule
\end{tabular}
\end{table*}
As shown in Table~\ref{tab:ad-scorers}, Mahalanobis distance achieves the highest AUROC for six of the eight anomalous decay channels, including all four near-OOD channels. 
For far-OOD anomalies, the leading scorer varies by channel: Isolation Forest reaches an AUROC of approximately 0.95 for channel F1, whereas $k$-NN distance performs best for channel F2, with an AUROC of 0.83. 
Isolation Forest falls to 0.49 for channel F2, illustrating the variation in sensitivity across decay trees. 
The ensemble does not consistently outperform the individual scorers, and Recon-MSE yields AUROC values between 0.41 and 0.54. 
Overall, Mahalanobis distance provides the most consistent performance across these test sets.

\begin{figure*}[t]
\centering
\includegraphics[width=\textwidth]{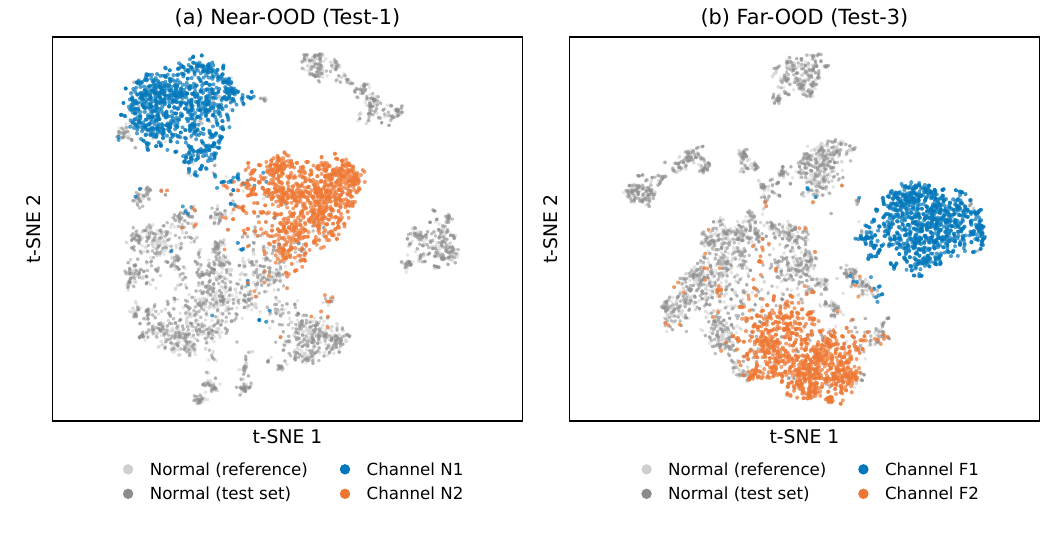}
\caption{
t-SNE projections of CLEAR event representations trained with the alignment--uniformity objective at $\lambda=10$ for (a) the near-OOD Test-1 and (b) the far-OOD Test-3 sets. ``Normal (reference)'' denotes events from the sample used to fit the anomaly scorer, and ``Normal (test set)'' denotes in-distribution events in the corresponding test set.}
\label{fig:ad-tsne}
\end{figure*}
%
%
\subsubsection{$\lambda$-sweep for Anomaly Detection}
\label{sec:ad-ood}
Based on the alignment-uniformity decomposition defined in Eq.~\ref{eq:alignment_uniformity}, we investigate how their relative weighting affects anomaly detection performance during pretraining. 
We pretrain five CLEAR variants with $\lambda \in \{0.1, 0.3, 1, 2, 10\}$. 
From the comparison in Table~\ref{tab:ad-scorers}, Mahalanobis distance is adopted as the anomaly scorer for the $\lambda$-sweep experiments, while the corresponding AUROC is used as the performance metric. 
We additionally include two reference baselines. 
The first baseline is an MPM-only model, in which the contrastive objective is disabled and only the masked modeling branch is retained. 
The second baseline employs the conventional InfoNCE objective instead of the explicitly decomposed alignment--uniformity loss. 
Together, these comparisons separate the contribution of contrastive learning itself from the effect of controlling the geometry of the representation space, thereby revealing how the balance between within-class compactness and global uniformity influences the separation of unseen decay topologies from known ones.

\begin{table}[t]
\centering
\caption{Mean AUROC obtained with Mahalanobis scoring under different pretraining. Each value is averaged over the two anomalous decay modes in the corresponding test set.}
\label{tab:ad-ood}
\small
\setlength{\tabcolsep}{7pt}
\renewcommand{\arraystretch}{1.15}
\begin{tabular}{lcccc}
\toprule
& \multicolumn{2}{c}{Near-OOD}
& \multicolumn{2}{c}{Far-OOD} \\
\cmidrule(lr){2-3}\cmidrule(lr){4-5}
Pretraining
& Test-1 & Test-2 & Test-3 & Test-4 \\
\midrule
MPM-only        & 0.76 & 0.68 & 0.73 & 0.60 \\
InfoNCE         & 0.77 & 0.64 & 0.84 & 0.77 \\
\midrule
$\lambda=0.1$   & 0.77 & 0.65 & 0.86 & 0.85 \\
$\lambda=0.3$   & 0.81 & 0.66 & \textbf{0.86} & 0.84 \\
$\lambda=1$     & 0.84 & 0.71 & 0.84 &  0.87 \\
$\lambda=2$     & 0.84 & 0.70 & 0.84 & 0.86 \\
$\lambda=10$    & \textbf{0.86} & \textbf{0.73} & 0.85 & \textbf{0.87} \\
\bottomrule
\end{tabular}
\end{table}

Table~\ref{tab:ad-ood} shows that contrastive pretraining improves far-OOD detection over MPM-only pretraining both for Test-3 and Test-4.
Within the alignment--uniformity sweep, increasing $\lambda$ generally improves near-OOD detection: the mean AUROC rises from 0.77 to 0.86 on Test-1 and from 0.65 to 0.73 on Test-2.
For far-OOD detection, performance remains relatively stable across the sweep, with AUROC values of 0.84--0.86 on Test-3 and 0.84–0.87 on Test-4.
\textcolor{black}{These results suggest that increasing the relative weight of uniformity is particularly beneficial for near-OOD detection, whereas its additional benefit for far-OOD detection is limited because performance is already high at small values of $\lambda$.}

\textcolor{black}{Figure~\ref{fig:ad-tsne} visualizes the representations obtained at $\lambda=10$ for one near-OOD and one far-OOD test set.
In both cases, the two anomalous decay channels occupy distinct regions that are largely separated from the normal events and from each other, although some overlap remains.
Particularly, the separation in Test-1 is notable because its anomalous channels share final-state particle compositions with pretraining data, which suggests that the learned representations capture information relevant to resonance state.}

%
%
\subsection{Scaling Behavior}
\label{sec:scaling}
To investigate the effect of pretraining data scale, we pretrain the same CLEAR architecture on 2M and 10M events and evaluate the resulting models using three downstream classification strategies: an MLP-2L probe with a frozen encoder, CA-only fine-tuning and end-to-end fine-tuning.
\begin{table}[t]
\centering
\caption{Effect of pretraining data scale on 100 class classification accuracy for a frozen-encoder MLP-2L probe and two fine-tuning strategies. The improvement $\Delta$ is reported in percentage points.}
\label{tab:scaling}
\small
\begin{tabular}{lccc}
\toprule
Events & MLP-2L & CA-Only & E2E \\
\midrule
2M  & 69.06\% & 69.85\% & 71.91\% \\
10M & 69.29\% & 72.49\% & \textbf{74.53\%} \\
\midrule
$\Delta$ & +0.23\% & +2.64\% & +2.62\% \\
\bottomrule
\end{tabular}
\end{table}

As shown in Table~\ref{tab:scaling}, increasing the pretraining sample from 2M to 10M events only have a small improvement with the encoder frozen: the MLP-2L accuracy increases from 69.06\% to 69.29\% with a gain of 0.23\%. 
Both fine-tuning strategies yield more substantial gains, with CA-only and end-to-end fine-tuning improving accuracy by 2.64\% to 2.62\%, respectively. 
The end-to-end fine-tuning strategy achieves the highest accuracy of 74.53\%.

These results indicate that, the benefits of additional pretraining data are more pronounced following fine-tuning than with a frozen encoder at fixed model capacity.
The similar gains of approximately 2\% from CA-only and end-to-end fine-tuning, compared with the small improvement with a frozen encoder, suggest that adapting the event-level aggregation helps realize the benefits of larger-scale pretraining without requiring updates to the full encoder. 
However, with only two pretraining data scales evaluated, these findings support a positive data-scaling trend rather than establish a quantitative scaling law.

%
%
\begin{figure*}[t]
\centering
\includegraphics[width=\textwidth]{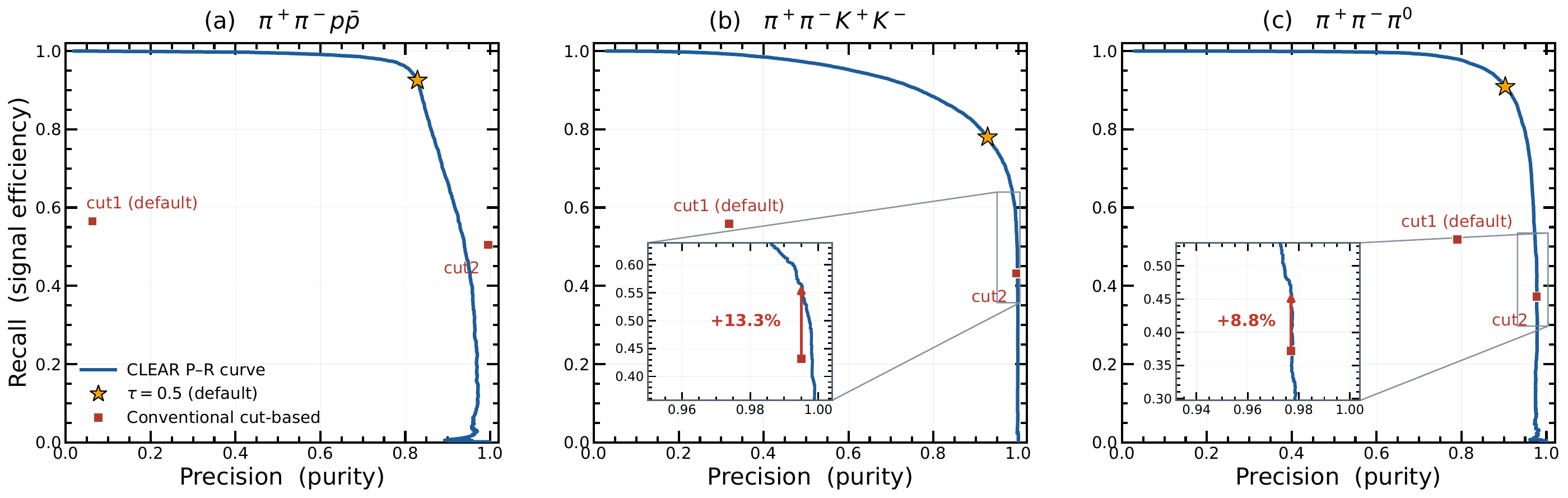}
\caption{
Precision--recall curves for CLEAR and conventional cut-based selections in (a)~$J/\psi\to\pi^{+}\pi^{-}p\bar{p}$, (b)~$J/\psi\to\pi^{+}\pi^{-}K^{+}K^{-}$ and (c)~$J/\psi\to\pi^{+}\pi^{-}\pi^{0}$.
Blue curves show CLEAR as its classification threshold varies, gold stars mark $\tau=0.5$, and red squares mark the conventional cut1 and cut2 selections.
At the purity of cut2, the insets show CLEAR efficiency gains of 13.3\% and 8.8\% in (b) and (c), respectively, but no matched-purity comparison is available in (a).
}
\label{fig:precision-recall}
\end{figure*}

\subsection{Comparison with Conventional Cut-Based Analyses}
\label{sec:conventional}
Conventional cut-based analyses at BESIII typically construct a dedicated selection pipeline for each exclusive decay channel, combining charged-track and EMC-shower selection, particle-identification requirements and kinematic fits~\cite{cut1cut2}.
Designing and optimizing these criteria relies heavily on expert knowledge and must generally be repeated for each newly investigated channel.
By contrast, CLEAR learns a general event-level representation from particle-level detector observables and can be adapted to different channels.

Conventional selections are essentially channel-specific binary decisions that distinguish a target signal from background.
Their performance therefore cannot be compared directly with the 100 decay channels classification accuracy reported in Sec.~\ref{sec:classification}.
To enable a consistent comparison, we formulate an exclusive final-state identification task for three representative decay channels: $J/\psi\to\pi^{+}\pi^{-}p\bar{p}$, $J/\psi\to\pi^{+}\pi^{-}K^{+}K^{-}$ and $J/\psi\to\pi^{+}\pi^{-}\pi^{0}$.
For each task, the target decay mode is treated as the signal, while all the other events in a held-out sample of $5\times10^{5}$ inclusive $J/\psi$ decays constitute the background.
The CLEAR decision threshold is varied to obtain the precision--recall curves shown in Fig.~\ref{fig:precision-recall}, where precision represents signal purity and recall represents signal efficiency.
\textcolor{black}{The conventional analyses provide two operating points for each channel, denoted here by cut1 and cut2~\cite{cut1cut2}}.
\textcolor{black}{For the $\pi^{+}\pi^{-}p\bar{p}$ and $\pi^{+}\pi^{-}K^{+}K^{-}$ channels, cut1 requires good charged tracks but does not impose particle identification or kinematic-fit requirements. 
Cut2 additionally applies particle identification and four-constraint (4C) kinematic fits, which enforce energy--momentum conservation, under the $\pi^{+}\pi^{-}K^{+}K^{-}$ and $\pi^{+}\pi^{-}p\bar{p}$ hypotheses. 
The signal hypothesis is required to have a smaller fit $\chi^{2}$ than either competing hypothesis.
For the $\pi^{+}\pi^{-}\pi^{0}$ channel, cut1 requires two good charged tracks and a $\pi^{0}$ candidate reconstructed from two photons. Cut2 further applies particle identification and tighter kinematic-fit requirements.}

At the default threshold of $\tau=0.5$, events with a CLEAR signal score above 0.5 are classified as signal.
As reported in Table~\ref{tab:vs_conventional}, CLEAR achieves $F_1$ scores of $87.4\%$, $84.7\%$, and $90.6\%$ for the three decay modes, respectively, outperforming both conventional selections in all cases.
Thus, CLEAR provides a more favorable balance between purity and efficiency without requiring the repeated design and optimization of a cascade of channel-specific selection criteria.

Because CLEAR at $\tau=0.5$ and cut2 operate at different purities, we additionally compare their signal efficiencies at matched purity.
For $J/\psi\to\pi^{+}\pi^{-}K^{+}K^{-}$, CLEAR achieves a signal efficiency of $56.5\%$ at the cut2 purity of $99.5\%$, compared with $43.2\%$ for cut2, an improvement of 13.3\%.
For $J/\psi\to\pi^{+}\pi^{-}\pi^{0}$, the corresponding efficiency increases from $37.2\%$ to $46.0\%$ at a matched purity of $97.7\%$, an improvement of 8.8\%.
CLEAR therefore achieves higher signal efficiency than the conventional selections at matched purity for these two decay channels.

For $J/\psi\to\pi^{+}\pi^{-}p\bar{p}$, however, CLEAR does not reach the $99.5\%$ purity obtained by cut2, revealing a current limitation of the model in the extreme high-purity regime.
Conventional analyses therefore remain indispensable for physical interpretability, the evaluation and control of systematic uncertainties, and the construction of high-purity signal regions.
Overall, CLEAR provides a better precision--recall balance and, where matched-purity comparisons are attainable, higher signal efficiency than the conventional selections.

\begin{table}[t]
\centering
\small
\setlength{\tabcolsep}{3pt}
\caption{Exclusive final-state identification on a held-out sample of $5\times10^{5}$ inclusive $J/\psi$ events. 
The matched-purity working point uses the purity of the tight conventional selection. 
``---'' indicates that this purity is not reached.}
\label{tab:vs_conventional}
\begin{tabular}{llccc}
\toprule
Final state & Method & Precision & Recall & $F_1$ \\
\midrule
$\pi^+\pi^-p\bar{p}$
 & Cut1 &  6.3\% & 56.5\% & 11.3\% \\
 & Cut2 & 99.5\% & 50.5\% & 67.0\% \\
 & CLEAR ($\tau=0.5$)   & 82.9\% & 92.5\% & \textbf{87.4\%} \\
 & CLEAR (matched) & --- & --- & --- \\
\midrule
$\pi^+\pi^-K^+K^-$
 & Cut1 & 31.9\% & 55.8\% & 40.6\% \\
 & Cut2 & 99.5\% & 43.2\% & 60.2\% \\
 & CLEAR ($\tau=0.5$)   & 92.8\% & 78.0\% & \textbf{84.7\%} \\
 & CLEAR (matched) & 99.5\% & 56.5\% & 72.0\% \\
\midrule
$\pi^+\pi^-\pi^0$
 & Cut1 & 79.0\% & 51.8\% & 62.6\% \\
 & Cut2 & 97.7\% & 37.2\% & 53.9\% \\
 & CLEAR ($\tau=0.5$)   & 90.3\% & 90.9\% & \textbf{90.6\%} \\
 & CLEAR (matched) & 97.7\% & 46.0\% & 62.6\% \\
\bottomrule
\end{tabular}
\end{table}

%
%
\section{Conclusion}
We introduced CLEAR, a foundation model for physics analysis at BESIII that combines masked modeling and contrastive learning through hybrid pretraining. 
The resulting representations support classification of 100 classes $J/\psi$ decay channels and anomaly detection.
Contrastive learning improves the clusterability and linear separability of the representation space, benefiting discriminative downstream tasks. 
We further investigated how alignment and uniformity shape this space and affect downstream performance. Decay channel classification favors a balance between alignment and uniformity. Anomaly detection benefits from greater uniformity, particularly for near-OOD decay channels.
Increasing the pretraining sample has limited gains with a frozen encoder but provides a more effective starting point for fine-tuning. Improving discrimination among channels with identical final-state compositions but different intermediate resonances remains a priority for the next version of CLEAR.
Compared with the evaluated cut-based selections, CLEAR achieves higher signal efficiency at matched purity for two studied channels, while providing a unified classification of the top 100 decay channels. 
Overall, CLEAR demonstrates the potential of hybrid pretraining to learn reusable event-level representations to do clusterization and classification for all the decay channels, thereby supporting a wide range of analysis tasks in HEP experiments.

\section*{Acknowledgments}
\textcolor{black}{This work was supported by the Strategic Priority Research Program of the Chinese Academy of Sciences under Grant No.~XDA0480600. 
The authors thank the BESIII Collaboration for its strong support, especially
for providing the inclusive $J/\psi$ Monte Carlo sample and the BOSS
analysis environment.
The authors also thank the IHEP Computing Center for its computational support.}

\section*{Data Availability}
\textcolor{black}{The BESIII inclusive $J/\psi$ Monte Carlo dataset is available through the National High Energy Physics Science Data Center~\cite{BESIIIInclusiveMC}.
The CLEAR source code is available on GitHub~\cite{CLEARCode}.}

\bibliography{apssamp}

\appendix
\clearpage
\onecolumngrid
\section{Single-particle masking}
\label{app:single-particle-masking}

Algorithm~\ref{alg:single-particle-masking} summarizes the single-particle masking strategy.
For each event, one valid particle is sampled uniformly and removed from the encoder input.
The function also records its original position and constructs the indices required to restore the particle ordering.
After encoding, a learnable mask token is appended to the decoder input and placed at the removed particle position.
The reconstruction loss is evaluated only for the selected particle and its valid features.

\begin{algorithm}
\caption{Single-particle masking}
\label{alg:single-particle-masking}
\begin{algorithmic}[1]
\Require Embedded tokens
$\mathbf{X}\in\mathbb{R}^{B\times(N+1)\times D}$,
with the \textsc{CLS} token at index $0$, and validity mask
$\mathbf{V}\in\{0,1\}^{B\times(N+1)}$
\Ensure Kept tokens $\widetilde{\mathbf{X}}$, prediction mask
$\mathbf{M}$, restoration indices $\mathbf{R}$, kept-token mask
$\widetilde{\mathbf{V}}$, and kept indices $\mathbf{K}$

\State $\mathbf{M}\gets\mathbf{0}_{B\times(N+1)}$

\For{$b=1$ to $B$}
    \State $\mathcal{I}_b\gets
    \{j\in\{1,\ldots,N\}\mid \mathbf{V}_{b,j}=1\}$
    \State Sample $m_b$ uniformly from $\mathcal{I}_b$

    \State $\mathbf{K}_b\gets
    [0]\mathbin{\|}[j\in\{1,\ldots,N\}:j\neq m_b]$
    \State $\widetilde{\mathbf{X}}_b
    \gets\mathbf{X}_{b,\mathbf{K}_b,:}$
    \State $\widetilde{\mathbf{V}}_b
    \gets\mathbf{V}_{b,\mathbf{K}_b}$
    \State $\mathbf{M}_{b,m_b}\gets1$

    \State $\boldsymbol{\pi}_b
    \gets\mathbf{K}_b\mathbin{\|}[m_b]$
    \For{$k=0$ to $N$}
        \State $\mathbf{R}_{b,\pi_{b,k}}\gets k$
    \EndFor
\EndFor

\State \Return
$\widetilde{\mathbf{X}},\mathbf{M},\mathbf{R},
\widetilde{\mathbf{V}},\mathbf{K}$
\end{algorithmic}
\end{algorithm}

\section{Class-wise classification performance}
\label{app:confusion-matrix}
\begin{figure*}[t]
\centering
\includegraphics[width=\textwidth]{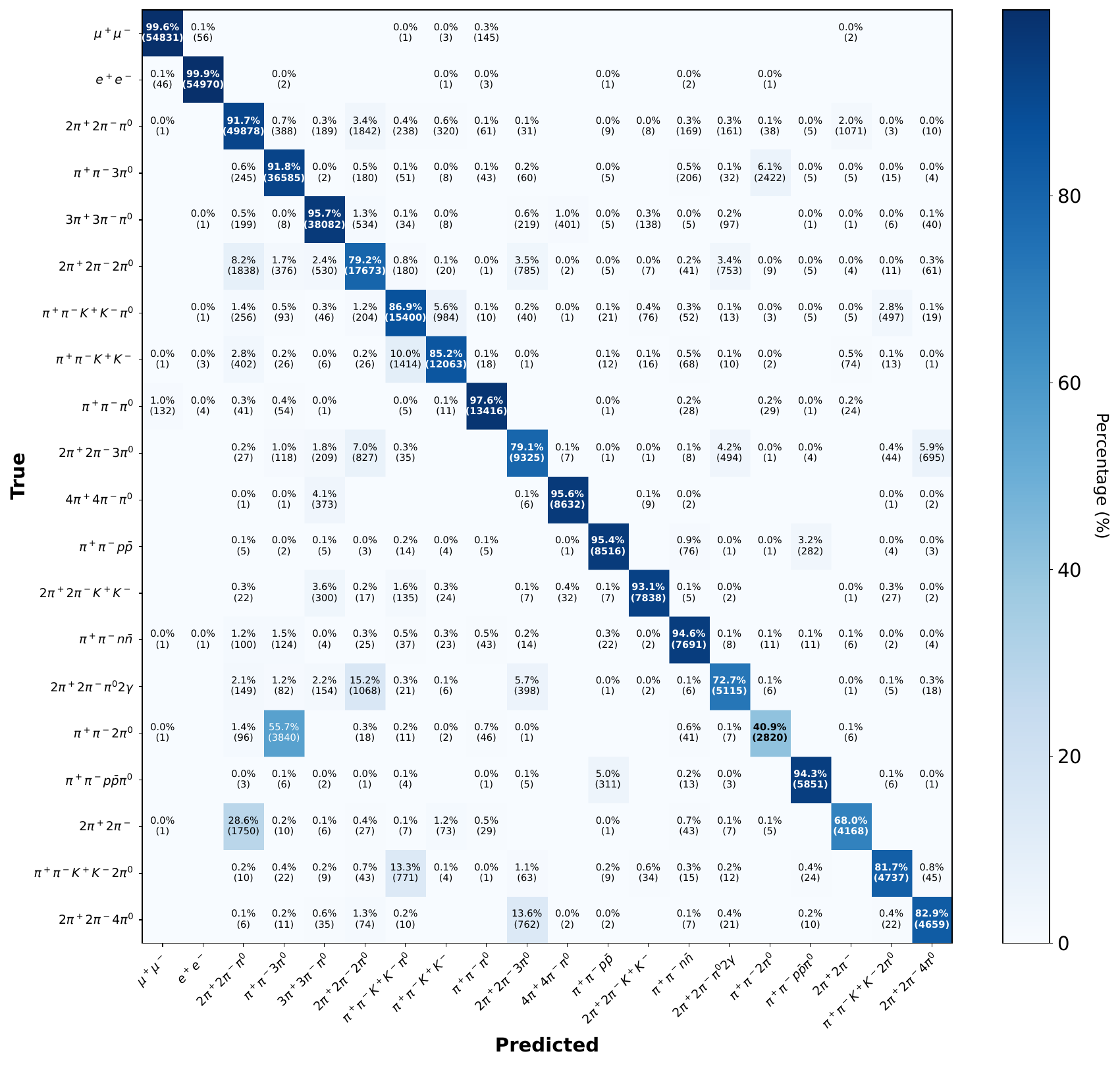}
\caption{
Row-normalized confusion matrix for the 20 most frequent decay modes in the 100-class classification task, obtained using a frozen InfoNCE-pretrained CLEAR encoder and a two-layer MLP classifier.
Percentages are shown with event counts in parentheses.
}
\label{fig:mlp-confusion-matrix}
\end{figure*}

To complement the aggregate classification metrics, we examine the class-wisepredictions of CLEAR using the confusion matrix shown in Fig.~\ref{fig:mlp-confusion-matrix}.
The encoder was pretrained with masked particle modeling and InfoNCE contrastive learning and subsequently kept frozen, while only the two-layer MLP classifier was trained on the 100-class downstream task.
For readability, the figure displays the 20 most frequent decay modes. 
Most classes exhibit a strong diagonal response, whereas the remaining misclassifications are concentrated among decay modes with similar final-state particle compositions.
Separating these modes requires the representation to encode more subtle kinematic and intermediate-resonance information.

\clearpage
\twocolumngrid

\end{document}